\documentclass[11pt,a4paper]{article}
\usepackage[left=1in, right=1in, top=1in, bottom=1in]{geometry}
\usepackage{pgfplots}
\pgfplotsset{width=7cm,compat=1.15}
\usepackage[toc,title,page]{appendix}
\usepackage{graphicx}
\usepackage{bm}
\newcommand*{\eqref}[1]{(\ref{#1})}
\usepackage[skip=8pt,font=scriptsize]{caption}
\usepackage{epstopdf}
\begin{document}
\begin{center}
{\large{A massive composite fermion from the three electron-like leptons in the bound state in the continuum}}\\

\vspace{.5cm}

A.I. Agafonov\\

\vspace{.5cm}

National Research Centre "Kurchatov Institute", Moscow 123182, Russia\\

Moscow Aviation Institute (National Research University), Moscow, 125993, Russia\\

\vspace{.5cm}

Agafonov\_AIV@nrcki.ru, aiagafonov7@gmail.com
\end{center}

\vspace{.5cm}


\vspace{.5cm}


\vspace{.5cm}

\begin{abstract}
In the bound state in continuum (BIC) the mass of a composite particle is greater than the total mass of its constituent particles. The BIC states that are stable due to confinement mechanisms, have been found in various fields but particle physics has evolved without analyzing these 
states.  The work's idea is to apply the BIC phenomenon to some elementary particles. The study focuses on a composite fermion of the three light leptons, $(e^{-}e^{+}e^{\pm})_{BIC}$. In the present theory the free electron and free positron is treated as different particles, each being characterized by the complete set of the Dirac plane waves. Using the Bethe-Salpeter equation, the wave function of the fermion with the mass of 207 electron masses was found numerically. The results obtained predict the particle radius equal to 65.9 Fm. The representation of muons as the composite fermions formed by the three electron-like leptons in the BIC states, is analyzed.
\end{abstract}

\vspace{.5cm}

\section{Introduction}
\par In the normal bound state the mass of the composite particle is less than the total mass of its constituent particles,
\begin{displaymath}
M<\sum_{i}m_{i}.
\end{displaymath}
It means that the energy eigenvalue of this state is negative.
\par A different situation is inherent in bound states in the continuum (BIC) for which the energy eigenvalues are positive.   
Respectively, for the composite particle mass is greater than the total mass of the constituent particles,
\begin{displaymath}
M>\sum_{i}m_{i}.
\end{displaymath}
\par BIC were first studied by von Neumann and Wigner in 1929 \cite{Neu} (see also \cite{Sti} with some extension of this work). The problem formulated in this work is follows. Let the particle moves in an external field that vanishes at infinity. Its behavior is described by the Schrödinger Hamiltonian. One must find the potential energy, for which the energy eigenvalue is positive, and the wave function corresponds to the spatially localized state. It turned out that there are many possible solutions for this problem \cite{Sti,Hsu,Sim}. These states are localized waves coexisting with the continuous spectrum of radiating waves that can carry energy away, but, nevertheless, are stable \cite{Hsu}. 
\par First proposed in quantum mechanics \cite{Neu}, BICs have then been found in various fields of physics including condensed matter and   and optic systems. The confinement mechanisms for these states are fundamentally different from those of the conventional bound states (see the review \cite{Hsu} and references therein). As far as we know, in quantum electrodynamics such states have not been supposed or studied previously. The idea of the work is to apply this phenomenon of BICs to some elementary particles. 
\par As a rule, in the non-relativistic limit, BICs have been investigated by using the Schrödinger Hamiltonian. Then, the spectral analysis of the real equation $H\psi=E\psi$ with $E>0$ showed that for any number of particles, the purely Coulomb Hamiltonians do not have BICs 
\cite{Sim}. However, the  eigenvalues of the BIC states are in the continuous spectrum of positive energies. Then the study of BICs can be carried out on the basis of integral equations, such as the homogeneous Lippmann-Schwinger integral equation, with the definition of the rule of pole bypassing \cite{Wut}. The kernels of integral equations and, respectively, the BIC wave functions become complex.
As a result, the spectral analysis should be provided for a system of two coupled integral equations.
\par In the present work we investigate the BIC state of a stable massive fermion composed of several light electron-like leptons. Moreover, it is assumed that the mass of composite particle is much larger than the total mass of the initial leptons, that is 
\begin{equation}\label{1}
M>>\sum_{i=1}^{N}m_{i}.
\end{equation}
For the composite fermions, the number of initial particles with the spin $\frac{1}{2}$ must be odd, starting with $N=3$.
\par For the case $N=3$, the structure of the positively charged composite fermion is determined by the two positron and the electron, that is $(e^{+}e^{+}e^{-})_{BIC}$. Similarly, the structure of the negatively charged particle corresponds to $(e^{-}e^{-}e^{+})_{BIC}$. The interaction between these particles is electromagnetic, and is given by the fine structure constant $\alpha=\frac{1}{137}$. 
\par Our consideration makes sense only for the case when the composite particle mass is approximately 207 times that of the electron.
This means that in this BIC state, all three leptons are relativistic. The question arises whether the electromagnetic interaction can
hold these particles together to form this BIC state.
\par The point is that there is the factor: 
\begin{equation}\label{2}
g^{(3)}=\frac{\alpha}{E-p_{1}-p_{2}-p_{3}+i\delta}. 
\end{equation}
in the integral equation for wave function, $F({\bf p}_{1},{\bf p}_{2},{\bf p}_{3}=-{\bf p}_{1}-{\bf p}_{2})$ in the momentum space. 
Here ${\bf p}_{1,2,3}$ are the particles' momenta. In the case the negative eigenvalues $E<0$, as is the case for ordinary bound states, 
$\delta$ can be omitted and no enhance of the interaction occurs. For BIC states the eigenvalues are positive, $E>0$, and the situation changes dramatically. Really, for the function \eqref{2} there is a continuous region of simple poles,
\begin{equation}\label{3}
p_{1}+p_{2}+\sqrt{p_{1}^2+p_{2}^2+2p_{1}p_{2}\cos\theta}=E,
\end{equation}
where $\theta$ is the angle between the momenta. This angle determines correlations in the particles' motions. 
This pole region is the 3D surface in the momentum space, as will be demonstrated below. The interaction between the particles is strongly enhanced on this 3D surface which represents the resonant region in momentum space, in which the interaction between particles becomes formally unlimited. 
Together with the correlations in particles motion, this leads to the confinement of these particles in the BIC state.
In other words, when the momenta of these three particles lie in this this 3D surface the resonance arises in the electromagnetic interaction between particles. This was the motivation for the present work.
\par Note that the resonance 3D surface is occurred only for the three-particle system. It is easy to verify that the resonant region is a volume for the four-particle system. The dimension of this resonance volume increases with increasing the number of interacting particles.  
\par Natural units ($\hbar =c=1$) will be used throughout.

\section{The equal-time Bethe-Salpeter equation}
As far as we know, up to now the closed form of the two-fermion interaction function was not obtained (see \cite{Bet}).
The summation of various diagrams that take into account the first contribution to this function leads to the well-known ladder approximation.
It is believed that in the case of the electromagnetic interaction the Bethe-Salpeter equation in the ladder approximation is an effective and unique tool allowing us to study bound states in the framework of quantum electrodynamics \cite{Nak,Efi,Wic,Cut,Sap}.
\par Various approaches have been developed for three-body problems \cite{Pie,Mar,Kar,Yam,Sav}. However, like many cases of the two-body problem, the three-body equations do not admit often the transition to the Schrödinger equation in the nonrelativistic limit. But it is important for us that this nonrelativistic limit takes place. Otherwise, it is difficult to understand whether the quantum state found is normal one.
\par  We treat the electron and positron as ordinary, different particles, each being characterized by the complete set of the Dirac plane 
waves. The rationale for this consideration is given in the work \cite{Aga}. We will show that this is a very important point so that the predicted decay channels of the three-particle system do not contradict the experimentally observed ones.
The lowest order of the pair interactions in the system will be taken into account, and the retardation of the interaction will be neglected. 
\par The three-particle system $(e^{-}e^{+}e^{\pm})_{BIC}$ contains two fermions which are identical and satisfy the exclusion principle. It should be taken into account when calculating the two-particle free propagator. We will assume that the pair of identical particles is in a singlet state in which the electronic states differ only in their spins. This allow us to simplify the expression for this two-particle propagation amplitude and, ultimately, to simplify the intermediate formulas. In our approach, the free three-particle propagator can be written as:
\begin{displaymath}
G=\sum_{{\bf p}_{1},{\bf p}_{2},{\bf p}_{3}}
\frac{1}{8\varepsilon_{{\bf p}_{1}}\varepsilon_{{\bf p}_{2}}\varepsilon_{{\bf p}_{3}}}
e^{i{\bf p}_{1}({\bf r}_{1}-{\bf r}_{1}^{\prime})+i{\bf p}_{2}({\bf r}_{2}-{\bf r}_{2}^{\prime})+
i{\bf p}_{3}({\bf r}_{3}-{\bf r}_{3}^{\prime})}
\Bigl[ \Lambda_{1}^{+}({\bf p}_{1})e^{-i\varepsilon_{{\bf p}_{1}}(t-t^{\prime})}+
\Lambda_{1}^{-}({\bf p}_{1})e^{+i\varepsilon_{{\bf p}_{1}}(t-t^{\prime})}
\Bigr] 
\end{displaymath}
\begin{equation}\label{4}
\Bigl[ \Lambda_{2}^{+}({\bf p}_{2})e^{-i\varepsilon_{{\bf p}_{2}}(t-t^{\prime})}+
\Lambda_{2}^{-}({\bf p}_{2})e^{+i\varepsilon_{{\bf p}_{2}}(t-t^{\prime})}
\Bigr]
\Bigl[ \Lambda_{3}^{+}({\bf p}_{3})e^{-i\varepsilon_{{\bf p}_{3}}(t-t^{\prime})}+
\Lambda_{3}^{-}({\bf p}_{3})e^{+i\varepsilon_{{\bf p}_{3}}(t-t^{\prime})}
\Bigr]
\end{equation}
Here 
\begin{equation}\label{5}
\Lambda_{i}^{\pm}({\bf p}_{i})=\varepsilon_{{\bf p}_{i}}\pm {\bm \alpha}_{i}{\bf p}_{i}
\pm m\beta_{i}
\end{equation}
$m$ is the electron mass, ${\bf p}_{i}$ and $\varepsilon_{{\bf p}_{i}}$ are the momentum and the energy of the $i-$ particle, $i=1,2,3$ and the ${\bm \alpha}$ and $\beta$ matrices are taken in the standard representation.
\par The Fourier transformation of \eqref{4} is given by:
\begin{displaymath}
G({\bf r}_{1}-{\bf r}_{1}^{\prime},{\bf r}_{2}-{\bf r}_{2}^{\prime},{\bf r}_{3}-{\bf r}_{3}^{\prime};E)=
\sum_{{\bf p}_{1},{\bf p}_{2},{\bf p}_{3}}
\frac{1}{8\varepsilon_{{\bf p}_{1}}\varepsilon_{{\bf p}_{2}}\varepsilon_{{\bf p}_{3}}}
e^{i{\bf p}_{1}({\bf r}_{1}-{\bf r}_{1}^{\prime})+i{\bf p}_{2}({\bf r}_{2}-{\bf r}_{2}^{\prime})+
i{\bf p}_{3}({\bf r}_{3}-{\bf r}_{3}^{\prime})}
\end{displaymath}
\begin{displaymath}
\Bigl[ \frac{\Lambda_{1}^{+}\Lambda_{2}^{+}\Lambda_{3}^{+}}{E-\varepsilon_{p_{1}}-\varepsilon_{p_{2}}-\varepsilon_{p_{3}}}+ 
\frac{\Lambda_{1}^{+}\Lambda_{2}^{+}\Lambda_{3}^{-}}{E-\varepsilon_{p_{1}}-\varepsilon_{p_{2}}+\varepsilon_{p_{3}}}+
\frac{\Lambda_{1}^{+}\Lambda_{2}^{-}\Lambda_{3}^{+}}{E-\varepsilon_{p_{1}}+\varepsilon_{p_{2}}-\varepsilon_{p_{3}}}+
\frac{\Lambda_{1}^{+}\Lambda_{2}^{-}\Lambda_{3}^{-}}{E-\varepsilon_{p_{1}}+\varepsilon_{p_{2}}+\varepsilon_{p_{3}}}+
\end{displaymath}
\begin{equation}\label{6}
\frac{\Lambda_{1}^{-}\Lambda_{2}^{+}\Lambda_{3}^{+}}{E+\varepsilon_{p_{1}}-\varepsilon_{p_{2}}-\varepsilon_{p_{3}}}+ 
\frac{\Lambda_{1}^{-}\Lambda_{2}^{+}\Lambda_{3}^{-}}{E+\varepsilon_{p_{1}}-\varepsilon_{p_{2}}+\varepsilon_{p_{3}}}+
\frac{\Lambda_{1}^{-}\Lambda_{2}^{-}\Lambda_{3}^{+}}{E+\varepsilon_{p_{1}}+\varepsilon_{p_{2}}-\varepsilon_{p_{3}}}+
\frac{\Lambda_{1}^{-}\Lambda_{2}^{-}\Lambda_{3}^{-}}{E+\varepsilon_{p_{1}}+\varepsilon_{p_{2}}+\varepsilon_{p_{3}}}\Bigr] 
\end{equation}
Here the value $E$ is understood as $E+i\delta$ where $E$ is the real energy of the stable composite particle and $\delta$ is $0^{+}$.
\par Along with the Coulomb interaction, interactions through the vector potential should be taken into account. Using \eqref{6}, the equal-time Bethe-Salpeter equation for the composite particle at rest is written as:
\begin{displaymath}
\psi({\bf r}_{1},{\bf r}_{2},{\bf r}_{3};E)=\alpha\int{d{\bf r}_{1}^{\prime}}\int{d{\bf r}_{2}^{\prime}}\int{d{\bf r}_{3}^{\prime}}
G({\bf r}_{1}-{\bf r}_{1}^{\prime},{\bf r}_{2}-{\bf r}_{2}^{\prime},{\bf r}_{3}-{\bf r}_{3}^{\prime};E).
\end{displaymath}
\begin{equation}\label{7}
\left(
\frac{1-{\bm \alpha}_{1}{\bm \alpha}_{2}}{\vert {\bf r}_{1}^{\prime}-{\bf r}_{2}^{\prime}\vert}-
\frac{1-{\bm \alpha}_{1}{\bm \alpha}_{3}}{\vert {\bf r}_{1}^{\prime}-{\bf r}_{3}^{\prime}\vert}-
\frac{1-{\bm \alpha}_{2}{\bm \alpha}_{3}}{\vert {\bf r}_{2}^{\prime}-{\bf r}_{3}^{\prime}\vert}
\right)\psi({\bf r}_{1}^{\prime},{\bf r}_{2}^{\prime},{\bf r}_{3}^{\prime};E)
\end{equation}
Here $\alpha$ is the fine structure constant.
\par In the nonrelativistic limit, $\vert E-3m \vert << m$ and $p_{i}<<m$. So, the first term in square brackets in Eq.\eqref{6} should only be taken into account. Then using \eqref{5} and \eqref{6}, one can see that Eq. \eqref{7} is reduced to the Schrödinger equation for the three-body problem.
\par For the massive composite fermion with the mass of $207m$, the particle's energy $\varepsilon_{{\bf p}_{i}}>>m$. Hence, in Eq. \eqref{5} the terms $m\beta_{i}$ can be omitted. Then we have:
\begin{equation}\label{8}
\Lambda_{i}^{\pm}({\bf p}_{i})=p_{i}\kappa_{i}^{\pm}
\end{equation}
with 
\begin{equation}\label{9}
\kappa_{i}^{\pm}=1{\pm}{\bf n}_{i}{\bm \alpha}_{i}
\end{equation}
and the unit vectors ${\bf n}_{i}={\bf p}_{i}/{p_{i}}$. As a result, the propagator \eqref{6} is reduced to the form: 
\begin{displaymath}
G({\bf r}_{1}-{\bf r}_{1}^{\prime},{\bf r}_{2}-{\bf r}_{2}^{\prime},{\bf r}_{3}-{\bf r}_{3}^{\prime};E)
=\frac{1}{8}\sum_{{\bf p}_{1},{\bf p}_{2},{\bf p}_{3}}
e^{i{\bf p}_{1}({\bf r}_{1}-{\bf r}_{1}^{\prime})+i{\bf p}_{2}({\bf r}_{2}-{\bf r}_{2}^{\prime})+
i{\bf p}_{3}({\bf r}_{3}-{\bf r}_{3}^{\prime})}
\end{displaymath}
\begin{displaymath}
\Bigl[ \frac{\kappa_{1}^{+}\kappa_{2}^{+}\kappa_{3}^{+}}{E-p_{1}-p_{2}-p_{3}}+ 
\frac{\kappa_{1}^{+}\kappa_{2}^{+}\kappa_{3}^{-}}{E-p_{1}-p_{2}+p_{3}}+
\frac{\kappa_{1}^{+}\kappa_{2}^{-}\kappa_{3}^{+}}{E-p_{1}+p_{2}-p_{3}}+
\frac{\kappa_{1}^{+}\kappa_{2}^{-}\kappa_{3}^{-}}{E-p_{1}+p_{2}+p_{3}}+
\end{displaymath}
\begin{equation}\label{10}
\frac{\kappa_{1}^{-}\kappa_{2}^{+}\kappa_{3}^{+}}{E+p_{1}-p_{2}-p_{3}}+ 
\frac{\kappa_{1}^{-}\kappa_{2}^{+}\kappa_{3}^{-}}{E+p_{1}-p_{2}+p_{3}}+
\frac{\kappa_{1}^{-}\kappa_{2}^{-}\kappa_{3}^{+}}{E+p_{1}+p_{2}-p_{3}}+
\frac{\kappa_{1}^{-}\kappa_{2}^{-}\kappa_{3}^{-}}{E-p_{1}+p_{2}+p_{3}}\Bigr] 
\end{equation}
For the state under investigation, this propagator \eqref{10} should be used in Eq. \eqref{7}.
\par We introduce the new variables:
\begin{equation}\label{11}
3{\bf R}={\bf r}_{1}+{\bf r}_{2}+{\bf r}_{3}, 
{\bf r}_{1}^{*}={\bf r}_{1}-{\bf r}_{3}, 
{\bf r}_{2}^{*}={\bf r}_{2}-{\bf r}_{3}.
\end{equation}
Here ${\bf r}_{1}$ and ${\bf r}_{2}$ are the radius-vectors of the both electrons, and ${\bf r}_{3}$ is the positron radius-vector.
According to \eqref{11}, the radius-vectors of both the electrons are counted from the positron position. Further, the superscript for the variables ${\bf r}_{1,2}^{*}$ is omitted. Using \eqref{11}, Eq. \eqref{7} with account for Eq. \eqref{10} takes the form:
\begin{displaymath}
\psi({\bf r}_{1},{\bf r}_{2};E)=\frac{\alpha}{8(2\pi)^6}\int{d{\bf r}_{1}^{\prime}}\int{d{\bf r}_{2}^{\prime}}
\int{d{\bf p}_{1}}\int{d{\bf p}_{2}}
e^{i{\bf p}_{1}({\bf r}_{1}-{\bf r}_{1}^{\prime})+i{\bf p}_{2}({\bf r}_{2}-{\bf r}_{2}^{\prime})}
\end{displaymath}
\begin{displaymath}
\Bigl[ \frac{\kappa_{1}^{+}\kappa_{2}^{+}\kappa_{3}^{+}}{E-p_{1}-p_{2}-p_{3}}+ 
\frac{\kappa_{1}^{+}\kappa_{2}^{+}\kappa_{3}^{-}}{E-p_{1}-p_{2}+p_{3}}+
\frac{\kappa_{1}^{+}\kappa_{2}^{-}\kappa_{3}^{+}}{E-p_{1}+p_{2}-p_{3}}+
\frac{\kappa_{1}^{+}\kappa_{2}^{-}\kappa_{3}^{-}}{E-p_{1}+p_{2}+p_{3}}+
\end{displaymath}
\begin{displaymath}
\frac{\kappa_{1}^{-}\kappa_{2}^{+}\kappa_{3}^{+}}{E+p_{1}-p_{2}-p_{3}}+ 
\frac{\kappa_{1}^{-}\kappa_{2}^{+}\kappa_{3}^{-}}{E+p_{1}-p_{2}+p_{3}}+
\frac{\kappa_{1}^{-}\kappa_{2}^{-}\kappa_{3}^{+}}{E+p_{1}+p_{2}-p_{3}}+
\frac{\kappa_{1}^{-}\kappa_{2}^{-}\kappa_{3}^{-}}{E+p_{1}+p_{2}+p_{3}}\Bigr] 
\end{displaymath}
\begin{equation}\label{12}
\left(
\frac{1-{\bm \alpha}_{1}{\bm \alpha}_{2}}{\vert {\bf r}_{1}^{\prime}-{\bf r}_{2}^{\prime}\vert}-
\frac{1-{\bm \alpha}_{1}{\bm \alpha}_{3}}{r_{1}^{\prime}}-
\frac{1-{\bm \alpha}_{2}{\bm \alpha}_{3}}{r_{2}^{\prime}}
\right)\psi({\bf r}_{1}^{\prime},{\bf r}_{2}^{\prime};E)
\end{equation}
Here $p_{3}=\vert {\bf p}_{1}+{\bf p}_{2}\vert$, and in the definition \eqref{9} the unit vector
\begin{equation}\label{13}
{\bf n}_{3}=-\frac{{\bf p}_{1}+{\bf p}_{2}}{\vert {\bf p}_{1}+{\bf p}_{2} \vert}.
\end{equation}
\par Note that even for the two-electron system the bispinor part of the wave function cannot be written \cite{Bet}. This difficulty is caused by the interaction between the particles. For the three-particle system Eq. \eqref{12} needs to be simplified. Below the following approximations are used. Eq. \eqref{12} contains the three terms $1-{\bm \alpha}_{i}{\bm \alpha}_{j\neq i}$. This is the relativistic generalization of the classical expression $1-{\bf v}_{i}{\bf v}_{j}$ that takes into account the two-particle interaction through the vector potential. Here ${\bf v}_{i=1,2,3}$ is the velocities of the particles. Indeed, the velocity operator is defined as ${\bf v}=i[H,{\bf r}]$. Substitution of the Dirac Hamiltonian as $H$ into this expression leads to ${\bf v}={\bm \alpha}$. For the BIC state these velocities differ little from the speed of light in vacuum. Using the relativistic relationship between the velocity of the particle and its momentum in the limit $p_{1,2}>>m$, we apply the replacement: 
\begin{equation}\label{14}
1-{\bm \alpha}_{i}{\bm \alpha}_{j\neq i} \to  1-{\bf n}_{i}{\bf n}_{j}, 
\end{equation}
where ${\bf n}_{i=1,2,3}$ are the unit vectors, ${\bf n}_{i}={\bf p}_{i}/{p_{i}}$. 
\par Now back to the definitions \eqref{9}. In our consideration,
\begin{equation}\label{15}
{\bf n}_{i}{\bm \alpha}_{i}=\frac{{\bf p}_{i}}{p_{i}}\frac{{\bf v}_{i}}{c}=1
\end{equation}
for $i=1,2,3$. Using \eqref{14}-\eqref{15} , we obtain that $\kappa_{i}^{+}=2$ and $\kappa_{i}^{-}=0$, and only the first term will remain 
in the square brackets in Eq. \eqref{12}.  
\par We assume that both the electrons are in the singlet state, and that the composite particle wave function can be presented 
the following form:
\begin{equation}\label{16}
\psi=F({\bf r}_{1},{\bf r}_{2})S,
\end{equation}  
where $S$ is the three-particle bispinor part of the wave function, and the function $F({\bf r}_{1},{\bf r}_{2})$ is symmetric with respect 
to the permutations ${\bf r}_{1}$ and ${\bf r}_{2}$. 
\par Taking into account Eqs.\eqref{13}-\eqref{16}, Eq. \eqref{12} is rewritten as:
\begin{displaymath}
F({\bf r}_{1},{\bf r}_{2})=\alpha \int{d{\bf r}_{1}^{\prime}}\int{d{\bf r}_{2}^{\prime}}
\int \frac{{d{\bf p}_{1}}}{(2\pi)^3}\int \frac{{d{\bf p}_{2}}}{(2\pi)^3}
e^{i{\bf p}_{1}({\bf r}_{1}-{\bf r}_{1}^{\prime})+i{\bf p}_{2}({\bf r}_{2}-{\bf r}_{2}^{\prime})}
\frac{1}{E-p_{1}-p_{2}-\vert {\bf p}_{1}+{\bf p}_{2}\vert+i\delta}
\end{displaymath}  
\begin{equation}\label{17}
\left(
\frac{1-{\bf n}_{1}{\bf n}_{2}}{\vert {\bf r}_{1}^{\prime}-{\bf r}_{2}^{\prime}\vert}-
\frac{1-{\bf n}_{1}{\bf n}_{3}}{r_{1}^{\prime}}-
\frac{1-{\bf n}_{2}{\bf n}_{3}}{r_{2}^{\prime}}
\right)
F({\bf r}_{1}^{\prime},{\bf r}_{2}^{\prime}),
\end{equation}
where ${\bf n}_{i}={\bf p}_{i}/{p_{i}}$. 
\par Solutions of the integral equation \eqref{17} must be sought in the basis of symmetric functions, 
$F({\bf r}_{1}, {\bf r}_{2}) = F({\bf r}_{2}, {\bf r}_{1})$.

\section{The momentum-space equation}
Using the representation
\begin{equation}\label{18}
\frac{1}{r}=\frac{2}{(2\pi)^2}\int \frac{d{\bf k}}{k^2}e^{i{\bf kr}},
\end{equation}
the Fourier-transformed Eq. \eqref{17} with \eqref{18} is given by:
\begin{displaymath} 
F({\bf p}_{1},{\bf p}_{2})=\frac{1}{2\pi^2}\frac{\alpha}{E-p_{1}-p_{2}-\vert {\bf p}_{1}+{\bf p}_{2} \vert+i\delta}
\Bigl[ (1-\frac{{\bf p}_{1}{\bf p}_{2}}{p_{1}p_{2}}) \int \frac{d{\bf k}}{k^2} F({\bf p}_{1}-{\bf k},{\bf p}_{2}+{\bf k})
\end{displaymath}
\begin{equation}\label{19}
-(1+\frac{p_{1}^2+{\bf p}_{1}{\bf p}_{2}}{p_{1}\vert {\bf p}_{1}+{\bf p}_{2} \vert}) 
\int \frac{d{\bf k}}{k^2} F({\bf p}_{1}-{\bf k},{\bf p}_{2})
-(1+\frac{p_{2}^2+{\bf p}_{1}{\bf p}_{2}}{p_{2}\vert {\bf p}_{1}+{\bf p}_{2} \vert}) 
 \int \frac{d{\bf k}}{k^2} F({\bf p}_{1},{\bf p}_{2}+{\bf k})
\Bigr].
\end{equation}
\par Eq. \eqref{19} is the homogenous integral equation for BIC. Here the complex function
\begin{equation}\label{20}
g^{(3)}=\frac{1}{E-p_{1}-p_{2}-\vert {\bf p}_{1}+{\bf p}_{2} \vert+i\delta} 
\end{equation}
leads to sharp increasing the interaction intensity in the three-particle system
when the momenta ${\bf p}_{1}$, ${\bf p}_{2}$ of the particles and the angle between them lie on the resonant 3D surfaces. 
\par In nonrelativistic case of the single particle motion with the mass $m$ in the external field, the similar function can be introduced,   
\begin{displaymath}
g^{(1)}=\frac{1}{E-\varepsilon_{p}+i\delta}.
\end{displaymath}
For the conventional bound states the energy eigenvalues are negative, $E-m<0$. As a result, the function $g^{(1)}$ has no poles, and 
$i\delta$ can be omitted. For BICs the energy of this localized state $E-m>0$ \cite{Hsu,Sti}. Then the function $g^{(1)}$ has a single 
simple pole, $\varepsilon_{p}=E$, and, hence, the infinitesimal $\delta$ is significant.
\par For the system of two interacting particles the respective function takes the form: 
\begin{displaymath}
g^{(2)}=\frac{1}{E-\varepsilon_{p_{1}}-\varepsilon_{p_{2}}+i\delta}. 
\end{displaymath}
In the nonrelativistic case the two-particle problem is reduced to single-particle one. Consequently, for the function $g^{(2)}$ we have the same situation as for $g^{(1)}$.
\par Usually, a complex function can have has a finite number or a infinite sequence of simple poles. But the function \eqref{20} has a continuous region of simple poles. This pole region is the 3D surface in the momentum space, as demonstrated in Fig. 1. The presented data correspond to the composite fermion with the mass $E=207m$. 
\begin{figure}[ht]
\centering
\includegraphics[width=16cm]{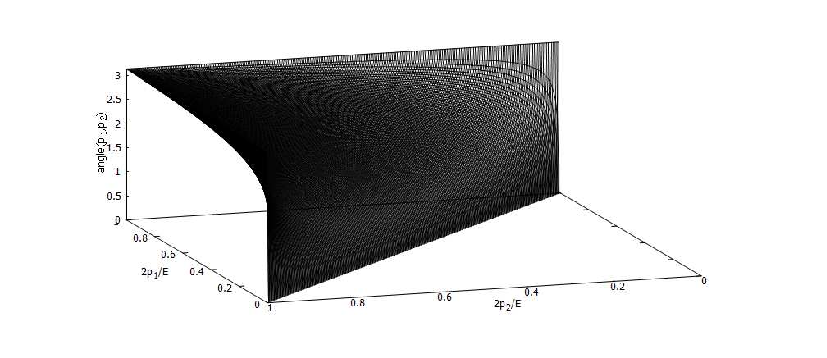}
\caption{
The 3D surface on which $E-p_{1}-p_{2}-\vert {\bf p}_{1}+{\bf p}_{2} \vert$ turns to zero. The composite fermion mass
$E=207m$.
\label{f1}}
\end{figure}
The interaction between the particles is strongly enhanced on this resonant 3D surface. Formally, this interaction becomes unlimited when $p_{1}+p_{2}+p_{3}=E$. One might say that the resonance arises in the electromagnetic interaction between particles on this momentum-space surface. Besides, these three particles forming the composite, are held relative to each other due to correlations in the particle motions. In Eq. \eqref{19} these correlations are represented by $(1-{\bf n}_{i}{\bf n}_{j})$. Both of these factors, namely the resonant surface and the correlations in the particle motions, can lead to the confinement of the particles in the BIC state.
\par The solution of the integral equation \eqref{19} is sought in the form:
\begin{equation}\label{21}
F({\bf p}_{1},{\bf p}_{2})=\sqrt{\frac{\delta}{\pi}}
\frac{Q({\bf p}_{1},{\bf p}_{2})}{E-p_{1}-p_{2}-\vert {\bf p}_{1}+{\bf p}_{2} \vert+i\delta}.
\end{equation}
From Eq. \eqref{21} we obtain the two-particle probability density:
\begin{equation}\label{22}
\vert F({\bf p}_{1},{\bf p}_{2}) \vert^{2}=\vert Q({\bf p}_{1},{\bf p}_{2}) \vert^{2} 
\delta(E-p_{1}-p_{2}-\vert {\bf p}_{1}+{\bf p}_{2} \vert).
\end{equation}
\par One can see that the two-particle probability density $\vert F({\bf p}_{1},{\bf p}_{2}) \vert^{2}$ is defined only on the resonant 3D surface presented in Fig. 1. Nevertheless, the function $Q({\bf p}_{1},{\bf p}_{2})$ can be defined in the whole momentum space.
\par Using Eq. \eqref{22}, the normalization condition of \eqref{21} is:
\begin{equation}\label{23}
\int d{\bf p}_{1}\int d{\bf p}_{2}
\vert Q({\bf p}_{1},{\bf p}_{2}) \vert^{2}
\delta(E-p_{1}-p_{2}-\vert {\bf p}_{1}+{\bf p}_{2} \vert)=1.
\end{equation}
Substituting \eqref{21} into \eqref{19}, we obtain: 
\begin{displaymath} 
Q({\bf p}_{1},{\bf p}_{2})=\frac{\alpha}{2\pi^2}\Bigl[
(1-\frac{{\bf p}_{1}}{p_{1}}\frac{{\bf p}_{2}}{p_{2}})\int \frac{d{\bf k}}{k^2} 
\frac{Q({\bf p}_{1}-{\bf k},{\bf p}_{2}+{\bf k})}{E-\vert{\bf p}_{1}-{\bf k}\vert-\vert{\bf p}_{2}+{\bf k}\vert-
\vert {\bf p}_{1}+{\bf p}_{2} \vert+i\delta}
\end{displaymath}
\begin{displaymath} 
-(1+\frac{p_{1}^2+{\bf p}_{1}{\bf p}_{2}}{p_{1}\vert {\bf p}_{1}+{\bf p}_{2} \vert})\int \frac{d{\bf k}}{k^2} 
\frac{Q({\bf p}_{1}-{\bf k},{\bf p}_{2})}{E-\vert{\bf p}_{1}-{\bf k}\vert-p_{2}-\vert {\bf p}_{1}+{\bf p}_{2}-{\bf k}\vert+i\delta}
\end{displaymath}
\begin{equation}\label{24}
-(1+\frac{p_{2}^2+{\bf p}_{1}{\bf p}_{2}}{p_{2}\vert {\bf p}_{1}+{\bf p}_{2}\vert})\int \frac{d{\bf k}}{k^2}
\frac{Q({\bf p}_{1},{\bf p}_{2}+{\bf k})}{E-p_{1}-\vert{\bf p}_{2}+{\bf k}\vert-\vert {\bf p}_{1}+{\bf p}_{2}+{\bf k}\vert+i\delta}\Bigr].
\end{equation}
Here normally $(x+i\delta)^{-1}=P(x^{-1})-i\pi\delta(x)$, where $P$ is the principal value of the function. It means that $P(x^{-1})=0$ 
for $x=0$. 

\section{The approximate solution of Eq. \eqref{24}}
Given the above results, it seems important to present the behavior of the composite fermion wave function and its characteristic sizes. 
With this aim, we assume that due to the correlations in the motion of the particles in BIC, the momenta of the two electrons, ${\bf p}_{1}$ and ${\bf p}_{2}$, have the same direction. Then, according to \eqref{24}, the electromagnetic interaction between the electrons vanishes since their Coulomb interaction is completely compensated by the interaction through the vector potential, $(1-{\bf n}_{1}{\bf n}_{2})=0$. Because 
${\bf p}_{3}=-({\bf p}_{1}+{\bf p}_{2})$, the momenta directions of all the three particles are uniquely defined. Hence, Eq. \eqref{24} is rewritten as:
\begin{equation}\label{25} 
Q(p_{1},p_{2})=-\frac{\alpha}{\pi^2}\int \frac{d{\bf k}}{k^2}\Bigl[ 
\frac{Q(\vert {\bf p}_{1}-{\bf k}\vert,p_{2})}{E-\vert{\bf p}_{1}-{\bf k}\vert-p_{2}-\vert {\bf p}_{3}+{\bf k}\vert+i\delta}
+\frac{Q(p_{1},\vert {\bf p}_{2}+{\bf k}\vert )}{E-p_{1}-\vert{\bf p}_{2}+{\bf k}\vert-\vert {\bf p}_{3}-{\bf k}\vert+i\delta}\Bigr].
\end{equation}
Solving Eq. \eqref{25} for $Q(p_{1},p_{2})$, we find the wave function \eqref{21}. In our case, the latter takes the form:
\begin{equation}\label{26}
F(p_{1}, p_{2})=\sqrt{\frac{\delta}{\pi}}
\frac{Q(p_{1},p_{2})}{E-2p_{1}-2p_{2}+i\delta}.
\end{equation}
\par From Eq. \eqref{26} we obtain the two-particle probability density:
\begin{equation}\label{27} 
\vert F(p_{1},p_{2}) \vert^{2}=\vert Q(p_{1},p_{2}) \vert^{2} \delta(2p_{1}+2p_{2}-E).
\end{equation}
Respectively, the normalization condition is:
\begin{equation}\label{28}
2^{3} \pi^{2} \int_{0}^{\frac{1}{2}E} p_{1}^{2} \Bigl(\frac{1}{2}E-p_{1}\Bigr)^2 \vert Q(p_{1},\frac{1}{2}E-p_{1}) \vert^{2}dp_{1}=1
\end{equation}
Note that the two-particle probability density and the normalization condition are defined only on the line 
\begin{equation}\label{29}
p_{1}+p_{2}=\frac{1}{2}E.
\end{equation}
This line \eqref{29} is the trace on the plane $(p_{1},p_{2})$ from the surface shown in Fig. 1. Nevertheless, the function $Q(p_{1},p_{2})$  is defined outside this line as well. 
\par In Eq. \eqref{25} the pole function has the representation: 
\begin{equation}\label{30} 
\frac{1}{E-\vert{\bf p}_{1}-{\bf k}\vert-p_{2}-\vert {\bf p}_{3}+{\bf k}\vert+i\delta}=
\frac{P}{E-\vert{\bf p}_{1}-{\bf k}\vert-p_{2}-\vert {\bf p}_{3}+{\bf k}\vert}-
i\pi \delta(E-\vert{\bf p}_{1}-{\bf k}\vert-p_{2}-\vert {\bf p}_{3}+{\bf k}\vert),
\end{equation}
where P means the principal part. 
On the right-hand side of Eq. \eqref{30}, the $\delta$-function defines the region of the resonant enhance of the interaction between particles. For \eqref{30} this region is given by the equation:
\begin{displaymath}
\sqrt{p_{1}^2+k^2-2p_{1}k\cos\theta}+p_{2}+\sqrt{(p_{1}+p_{2})^2+k^2-2(p_{1}+p_{2})k\cos\theta}=E, 
\end{displaymath}
where $\theta$ is the angle between the vectors ${\bf k}$ and ${\bf p}_{1,2}$. By definition of the principal part, inside this region we have:
\begin{displaymath}
\frac{P}{E-\vert{\bf p}_{1}-{\bf k}\vert-p_{2}-\vert {\bf p}_{1}+{\bf p}_{2}-{\bf k}\vert}=0.
\end{displaymath}
\par The resonant region is fundamentally important for the formation of the massive composite fermion with the mass $E=207m$. Therefore, in this resonance region, Eq.\eqref{25} takes the form:
\begin{displaymath} 
Q(p_{1}, p_{2})=\frac{i\alpha}{\pi} \int \frac{d{\bf k}}{k^2} 
Q(\vert {\bf p}_{1}-{\bf k} \vert, p_{2})\delta (E-\vert {\bf p}_{1}-{\bf k} \vert-p_{2}-\vert {\bf p}_{1}+{\bf p}_{2}-{\bf k} \vert)
\end{displaymath} 
\begin{equation}\label{31} 
+\frac{i\alpha}{\pi} \int \frac{d{\bf k}}{k^2} 
Q(p_{1},\vert {\bf p}_{2}-{\bf k}\vert)\delta(E- p_{1}-\vert{\bf p}_{2}-{\bf k}\vert-\vert{\bf p}_{1}+{\bf p}_{2}-{\bf k}\vert).
\end{equation}
In Eq. \eqref{31} we pass to the new variables: ${\bf k}_{1}={\bf k}-{\bf p}_{1}$ in the first integral, and
${\bf k}_{2}={\bf k}-{\bf p}_{2}$ in the second one. Then, omitting the subscripts of integration variables, this equation is rewritten as:
\begin{displaymath} 
Q(p_{1}, p_{2})=
\frac{i\alpha}{\pi}\int\frac{d{\bf k}}{({\bf k}+{\bf p}_{1})^2} Q(k, p_{2})\delta (E-k-p_{2}-\vert{\bf p}_{2}-{\bf k}\vert)
\end{displaymath} 
\begin{equation}\label{32} 
+\frac{i\alpha}{\pi}\int\frac{d{\bf k}}{({\bf k}+{\bf p}_{2})^2}Q(p_{1},k)\delta(E- p_{1}-k-\vert{\bf p}_{1}-{\bf k}\vert).
\end{equation}
Here $d{\bf k}=2\pi k^{2}dk dt$ with $\vert t \vert \leq 1$. 
\par An analysis of the equation \eqref{32} shows that the function $Q(p_{1},p_{2})$  is defined in the square 
$[0 \leq p_{1} \leq \frac{1}{2}E, 0 \leq p_{2} \leq \frac{1}{2}E]$ and $0 \leq k \leq \frac{1}{2}E$.    
\par Further, it is convenient to use the dimensionless variables: $x=2p_{1}/E$, $y=2p_{2}/E$, $z=2k/E$ and introduce the function: 
\begin{equation}\label{33}   
W(x,y)=\frac{1}{8}E^{5/2}Q(\frac{E}{2}x,\frac{E}{2}y).
\end{equation}
Then Eq. \eqref{32} is rewritten as:
\begin{displaymath}
W(x,y)=2i\alpha \int_{0}^{1} \int_{-1}^{+1} \frac{z^2dzdt}{z^2+x^2+2xzt}W(z,y)\delta\bigl(2-y-z-\sqrt{z^2+y^2-2yzt})\Bigr) 
\end{displaymath}
\begin{equation}\label{34}
+2i\alpha \int_{0}^{1} \int_{-1}^{+1} \frac{z^2dzdt}{z^2+y^2+2yzt}W(x,z)\delta\bigl(2-x-z-\sqrt{z^2+x^2-2xzt})\Bigr) 
\end{equation}
\par Details of our solution of the two-variable integral equation with the imaginary kernel are given in Appendix A.
\par Note that partition of $W$ into the imaginary and real parts is, in a sense, arbitrary since \eqref{34} is invariant under the phase transformation 
$W\to e^{i\phi}W$ with a constant $\phi$. Also, the function is symmetrical, $W(x,y)=W(y,x)$. Fig. 2 demonstrates the 
function $ReW(x,y)$.   
\begin{figure}[ht]
\centering
\includegraphics[width=16cm]{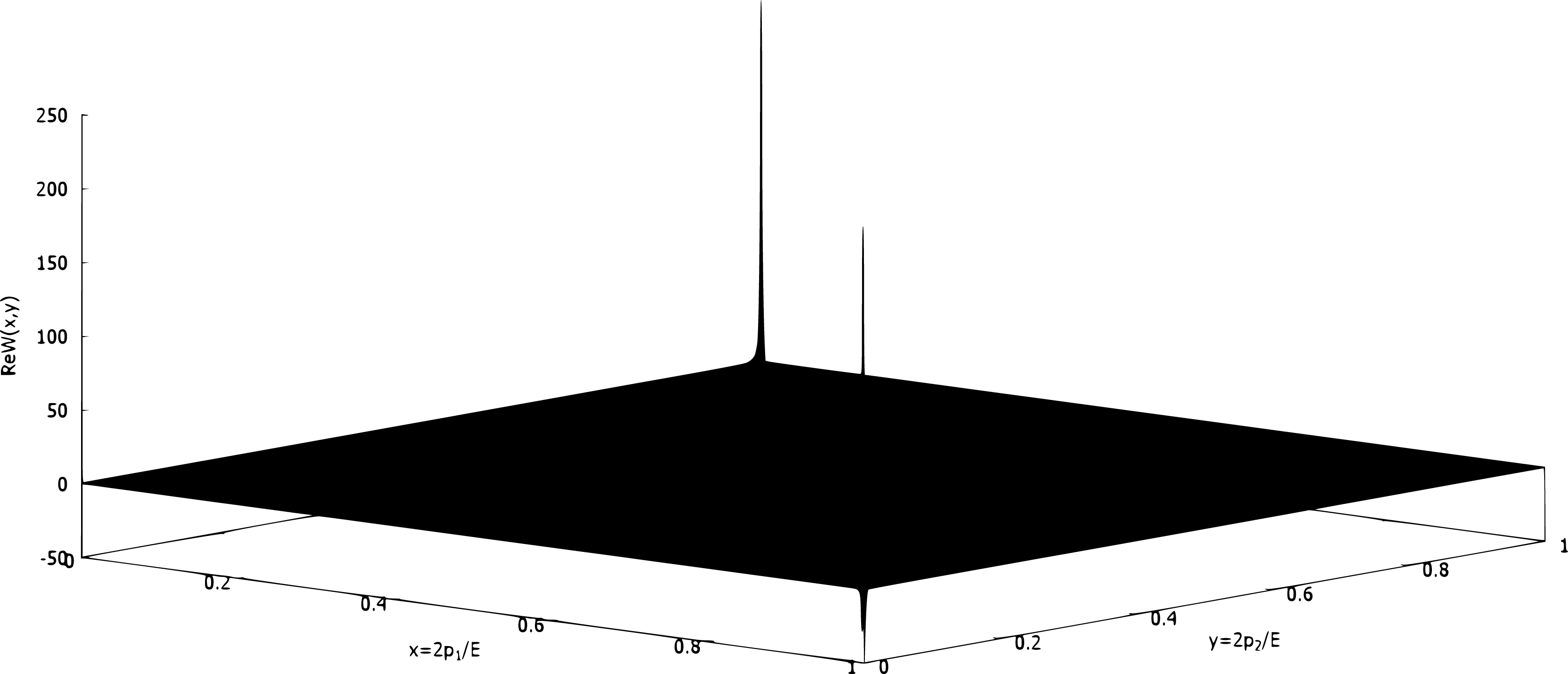}
\caption{The real part of the function $W(x,y)$ given by Eq. \eqref{33}. 
\label{f2}}
\end{figure}
In the presented scale grid, two identical narrow peaks are only clearly visible in the figure. One of them is located near the corner [0,1] of the function domain, and the other is in the symmetric corner [1,0]. Both of these peaks lie near the ends of the diagonal $x+y=1$ of the square. But the positions of the maxima of these peaks are off the diagonal.
\par The function $ImW$ has similar two peaks as well. They have the same shape and location in the function domain, but differ in sign from those presented in Fig. 2. The amplitude of these peaks for $ImW$ is about 270 that is somewhat larger than for the peaks shown in Fig. 2. These features are caused by the logarithmic singularity in Eqs. \eqref{A3} and \eqref{A4}. 
\par The key properties of the complex function $W$ can be described by the projection of the function on a plane. This projection plane contains the diagonal $x+y=1$ and is perpendicular to the plane $(x,y)$.  In order to present these features, we slightly change the graphical grid: $x,y \in [0.005,0.995]$. 
Then this projection for the function $ReW$ is represented in Fig. 3. One can see only the tails of the main peaks in the figure. The behavior of the function 
\begin{figure}[ht]
\centering
\includegraphics[width=16cm]{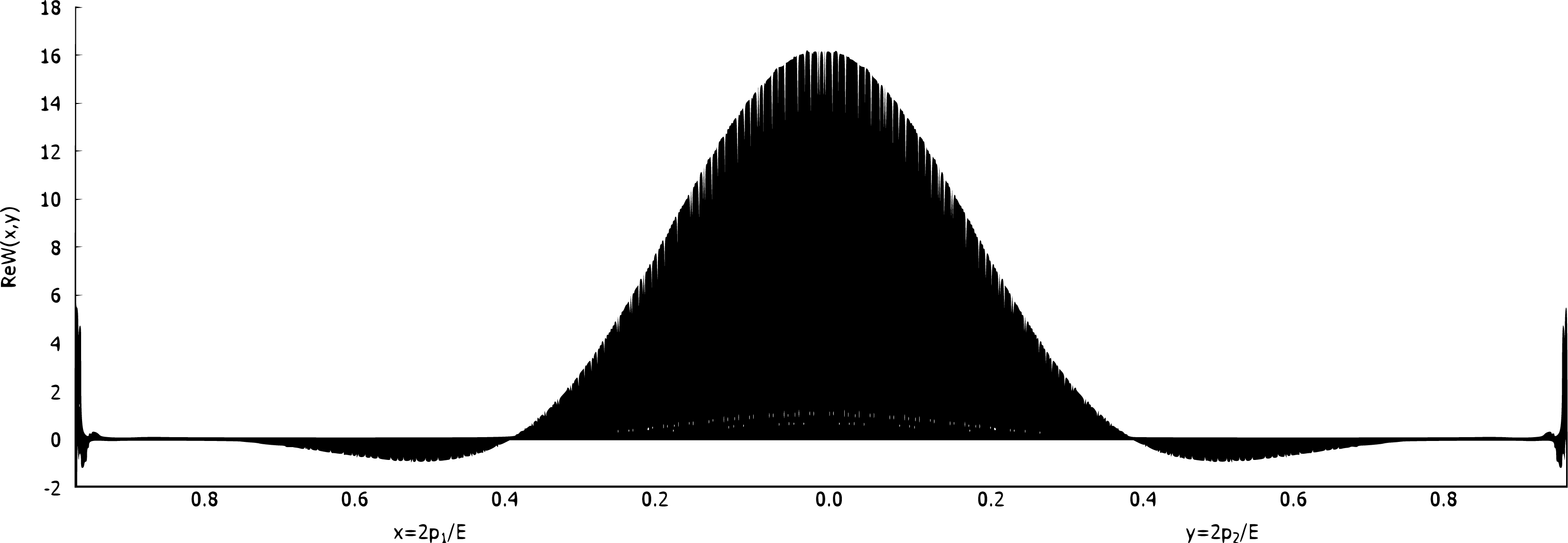}
\caption{The projection of $ReW$ on the plane that contains the diagonal $x+y=1$ and is perpendicular to the plane (x,y).
The graphical grid: $x,y \in [.005,0.995]$.
\label{f3}}
\end{figure}
Behavior of $W$ near this diagonal is important. It determines the normalization of the wave function in the BIC state of this three-particle system. This statement follows from Eqs. \eqref{27} and \eqref{27}. As shown in Fig. 3, the function $ReW$ has a central wide peak with two wings on either side with opposite sign to the main peak.The projection of the function $ImW$ also has a similar structure. But the peaks differ in sign from those presented in Fig. 3. 
\par The peaks shown in Fig. 3 have a rather complex structure in the transverse plane to the diagonal.
This structure of $ReW$  near the center of the diagonal is presented in Fig. 4. This structure could be described as peak-dip-peak. Exactly such structure is also characteristic of the function $ImW$.
\begin{figure}[ht]
\centering
\includegraphics[width=16cm]{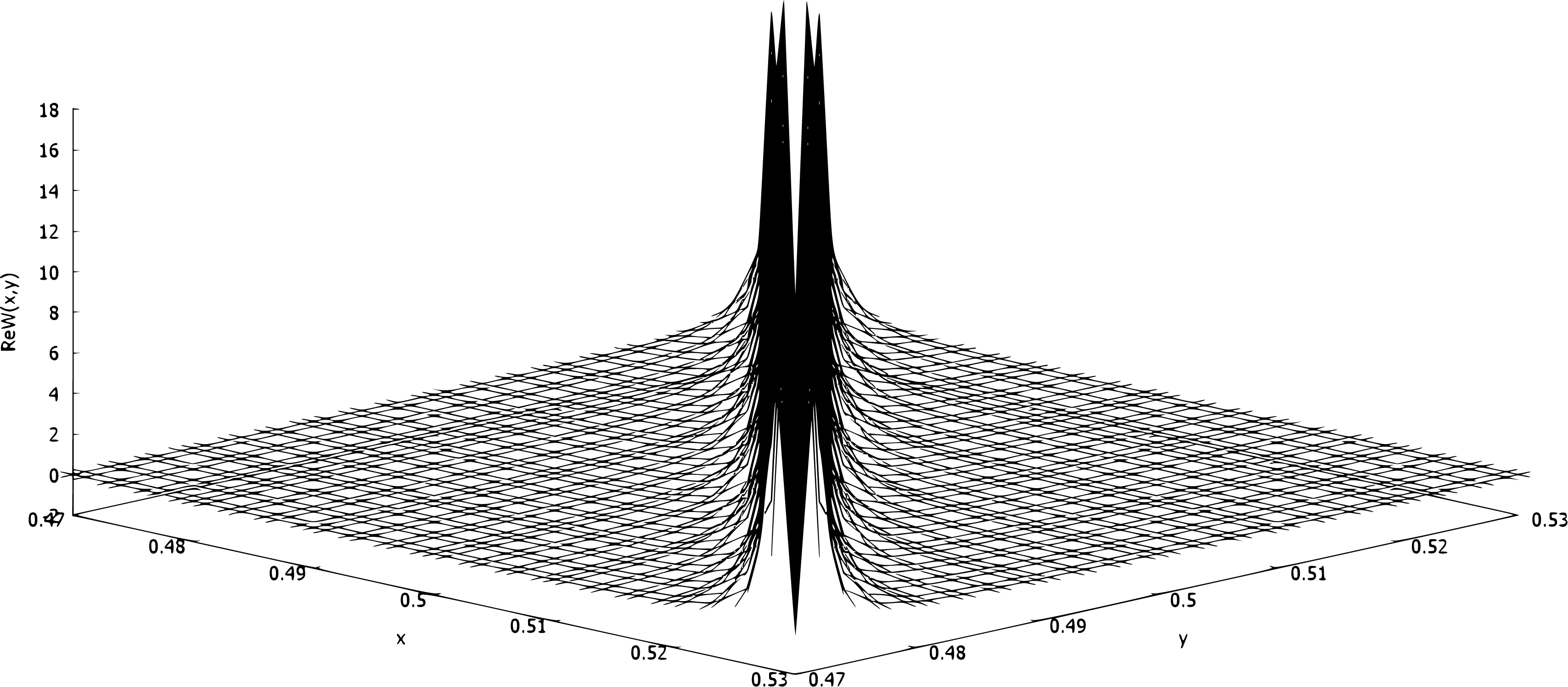}
\caption{The structure of $ReW$ near the center of the diagonal.
\label{f4}}
\end{figure}
\par When describing physical processes, the values of $W$ can be important in the entire domain of the function.
However, according to \eqref{27} and \eqref{28}, the normalization of the wave function of the composite fermion in the BIC state is determined only by the behavior 
of the complex function on the diagonal of the square. The functions $ReW(x,1-x)$ and $ImW(x,1-x)$ are
represented in Fig. 5.
\begin{figure}[ht]
\centering
\includegraphics[width=16cm]{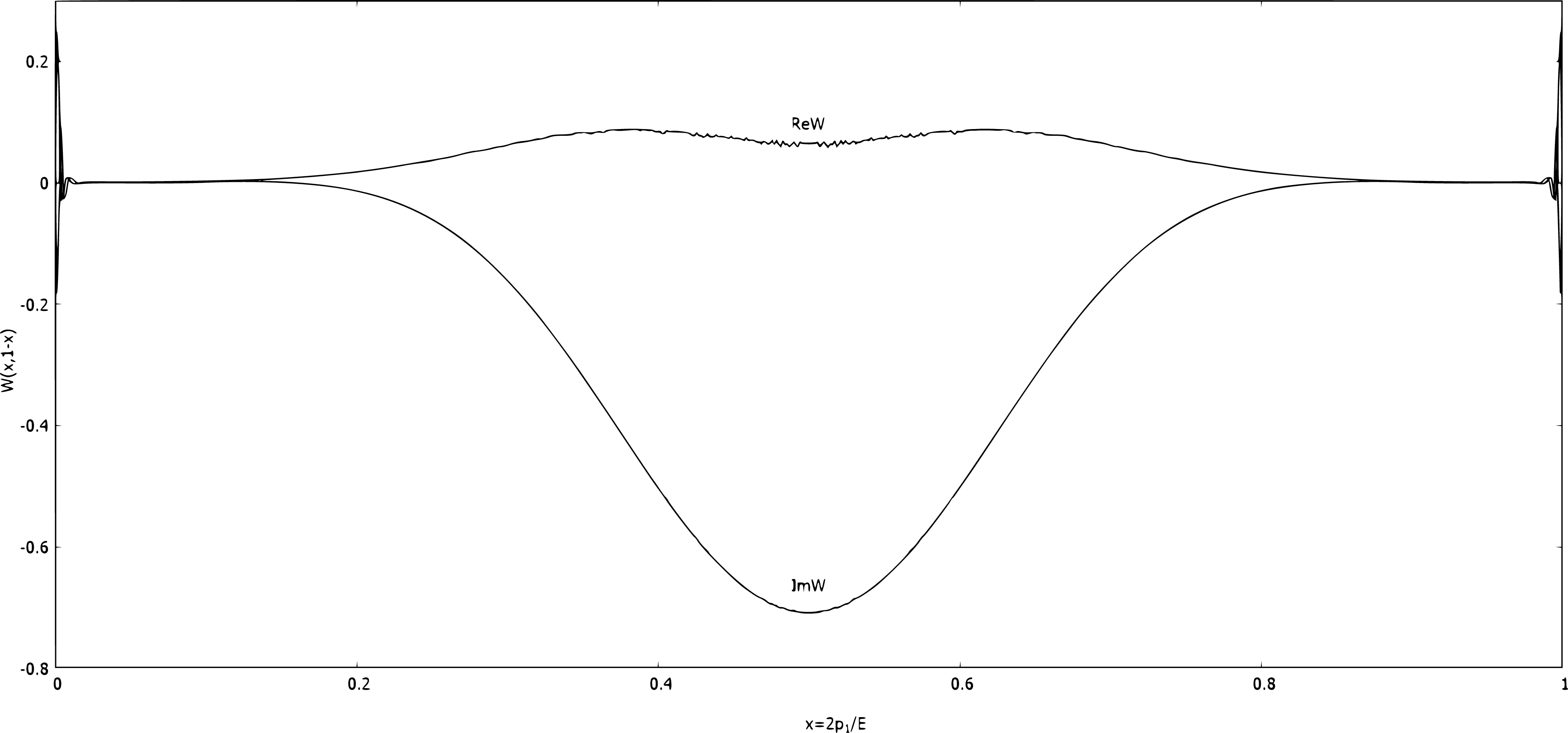}
\caption{The behavior of the functions $ReW(x,1-x)$ and $ImW(x,1-x)$ on the diagonal $x+y=1$.
\label{f5}}
\end{figure}
It can be seen that corner peaks on the ends of the diagonal do not make a significant contribution to the normalization. It is mainly determined by the behavior of the functions shown in Fig. 5, in the central part of the diagonal.
\par The presented functions correspond to the normalized wave function in the momentum space for the compound fermion. According to Eq. \eqref{B9}, the same functions,
$ReW(x,1-x)$ and $ImW(x,1-x)$, define the coordinate-space wave function of the fermion.  
Procedure for calculation of the coordinate-space wave function is presented in Appendix B. Figs. 6 and 7 show the real and imaginary parts of the wave function 
\eqref{B2}, respectively. Although both of these functions have similar behavior, both of them should be shown for the sake of completeness.
\begin{figure}[ht]
\centering
\includegraphics[width=16cm]{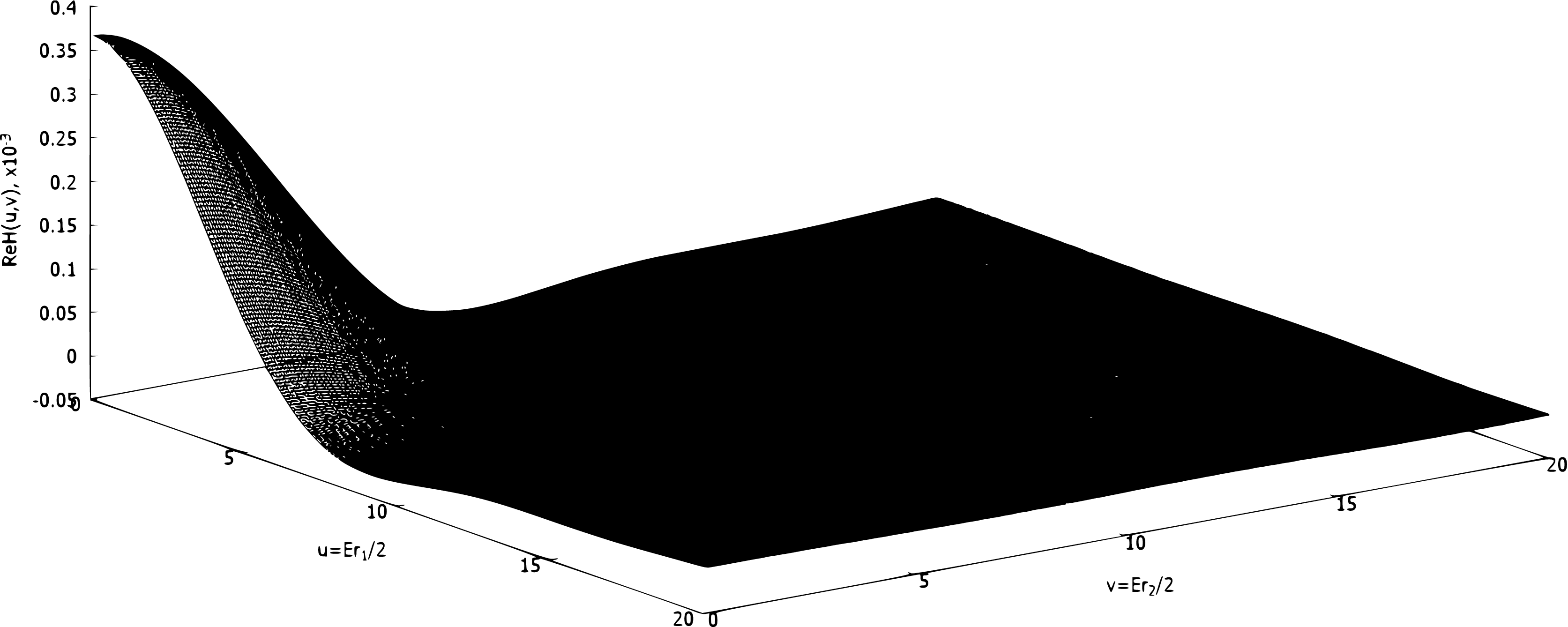}
\caption{
The real part of the wave function $H(u,v)$.
\label{f6}}
\end{figure}
The results presented in Figs. 6 and 7 correspond to the normalized wave function (see \eqref{B10}).
Obviously, the wave function has a symmetric maximum at $r_{1}=r_{2}=0$. Then, the composite fermion wave function shows damped oscillations with increasing 
$u=\frac{Er_{1}}{2}$ and $u=\frac{Er_{2}}{2}$.
\begin{figure}[ht]
\centering
\includegraphics[width=16cm]{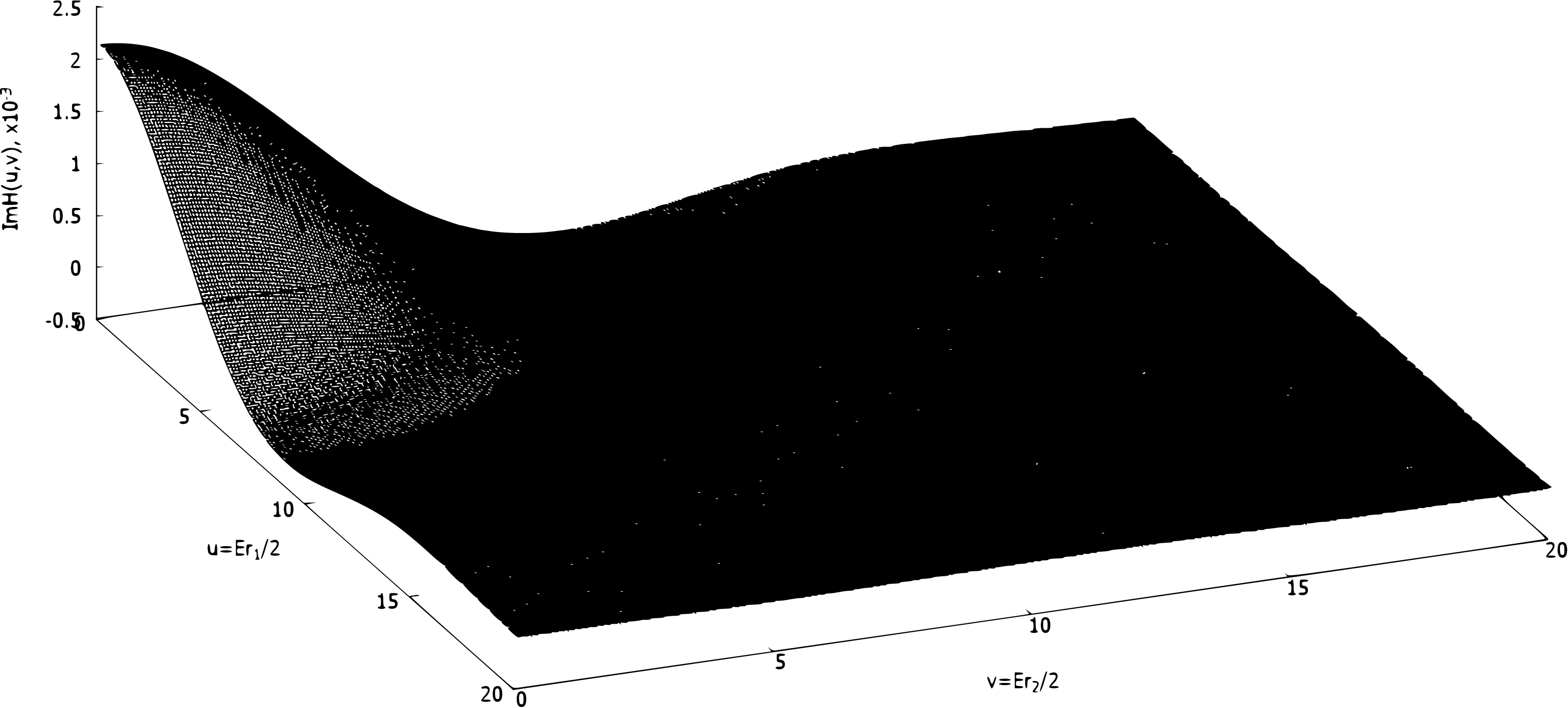}
\caption{
The imaginary part of the wave function $H(u,v)$.
\label{f7}}
\end{figure}
\par The usual characteristic of the wave function is its average radius. Its calculation was carried out according to Eq. \eqref{B11}.
It was found that for the results presented in Figs. 6 and 7 the average dimensionless radius is 
\begin{displaymath}
u_{av}=18.63
\end{displaymath}
In the usual units we have 
\begin{equation}\label{35}  
r_{av}=u_{av}\frac{2\hbar c}{E}
\end{equation}
For the fermion mass $E=207m$ we have $R=65.9$Fm. 
\par Below, in the next Section, we argue that the phenomenon of BIC for many-body systems may be very useful for understanding some problems in particle physics.

\section{Discussion and Conclusion}
Among the fundamental particles \cite{Rev}, particular attention is drawn to the particle from the lepton family, namely the muon which is known to be unstable. The fact that an unstable particle is accepted as the fundamental one is, in our opinion, strange but there is no other place for the muon in the Standard Model. 
\par Currently, the muons are considered as structureless \cite{Gor}. The muon mass is 207 times the electron mass, and its lifetime of approximately 2.2 $\mu$s. It is generally accepted that the muon decays into the electron, muon neutrino and electron antineutrino. As we understand, such the muon decay reaction is rather the theoretical scheme. The fact that the decay products are more than two particles, follows from conservation laws. However, except for the electron, it is hardly possible to detect experimentally both neutrinos during the decay \cite{Fray}. Besides, 
as far as we know, the decay channel $\mu^{-}\to e^{-}{\tilde\nu}_{e}{\nu}_{\mu}$ that is generally recognized as the main, is experimentally separated from the channel $\mu^{-}\to e^{-}{\tilde \nu}_{e}{\nu}_{\mu}\gamma$ only for photon energies greater than about 20 electron masses. Even the presence of photons with the lower energies is not excluded in the main channel 
(see \cite{Kun} and reference therein). 
\par The problem of interest to us is the following: whether the muon can be considered as the massive composite fermion formed by three charged electron-like leptons in the BIC state? The structure of the positive charged muon is represented as $\mu^{+}=(e^{+}e^{+}e^{-})_{BIC}$. The structure of the negatively charged muon is corresponded to $\mu^{-}=(e^{-}e^{-}e^{+})_{BIC}$. Accordingly, the results obtained above, are used to describe the muon wave function.
\par On this way, the following objections may arise. With hope to discover new physics, experiments have been conducted to search for decay processes, which are very suppressed or forbidden in the framework of the Standard Model. In this regard, many groups (see \cite{MEG,Bal,Fra} and references therein) have carried out experimental searches for the following muon decay channels:
\begin{equation}\label{36}
\mu^{\pm} \to e^{-}e^{+} e^{\pm}
\end{equation}
and
\begin{equation}\label{37}
\mu^{\pm} \to e^{\pm}\gamma.
\end{equation}
These reactions \eqref{36} and \eqref{37} are not prohibited by the laws of conservation of momentum and energy. However, the searches for these reactions are still unsuccessful up to now. Only very low values of the upper bounds  of these reactions were experimentally obtained.
Therefore, the lepton charge conservation law was formulated. This law is one of the foundations of the Standard Model of elementary particle physics. The reasons for the conservation of the lepton number are still unknown.
\par With respect to the reaction \eqref{36}, its prohibition is, generally speaking, not required lepton number conservation, since 
the states of the composite particles $\mu^{\pm}=(e^{+}e^{-}e^{\pm})$ are considered to be the BIC state. These states are localized waves which are stable and not-decaying. Due to the confinement mechanism the energy eigenvalue in BIC is real and positive, as in Eqs. \eqref{17} and \eqref{24}. So, the decay \eqref{36} does not occur for the composite fermion.  Therefore, according to the BIC theory, there is no point in searching for \eqref{36}. 
\par The main difficulty for our consideration is related to the decay channel \eqref{37}. In accordance with the central particle-antiparticle concept in QED \cite{Fey}, the conventional annihilation reaction between the electron and the positron which is considered as the antiparticle of the electron, is: 
\begin{equation}\label{38}
e^{-}e^{+} \to 2\gamma
\end{equation}
for the singlet electron-positron pair. In this process nothing remains of the electron and positron, and the reaction products are just a few photons \cite{Fey,Ore,Ber}. If the reaction \eqref{38} really takes place in nature, then the reaction \eqref{37} must also take place for the composite particles 
$(e^{+}e^{-}e^{\pm})_{BIC}$. In this case, our description of the muon structure is unacceptable. 
\par Thus, our consideration of the muons as $\mu^{\pm}=(e^{+}e^{-}e^{\pm})_{BIC}$ can only make sense if the annihilation reaction \eqref{38} is prohibited for certain reasons. Before discussing them, the physical rationale for \eqref{38} should be analyzed.
\par During the development of the QED theory, Feynman discussed two different free-fermion propagators for the Dirac equation \cite{Fey,Wei}:
\begin{equation}\label{39}  
K_{+}(2,1)=\sum_{\bf p}\psi_{p}(2)\bar{\psi }_{p}(1)\theta (t_{2} -t_{1} )-\sum_{\bf p}\psi_{-p} (2)\bar{\psi }_{-p}(1)  
\theta (t_{1}-t_{2})                          
\end{equation} 
and 
\begin{equation}\label{40}  
K_{-}(2,1)=\sum_{\bf p}\left(\psi_{p} (2)\bar{\psi}_{p} (1)+\psi_{-p} (2)\bar{\psi}_{-p} (1)\right) \theta (t_{2}-t_{1}).
\end{equation} 
\noindent Here $\psi_{\pm p}$ is the Dirac plane wave representing the state of the free particle with energy 
$\pm \varepsilon_{\bf p}$, respectively, and $\bar{\psi }_{p}$  the Dirac conjugate wave function.
\par The propagator \eqref{39} was originally intended for the hole theory of the positron and, then, was substantiated with the assumption of the fermion-antifermion symmetry in nature. The modern description of the electron-positron field is based on the use of \eqref{39} \cite{Fey,Ber}.
The contribution to $K_{+}(2,1)$ for $t_{2}>t_{1}$ is due to the electron terms, and that for $t_{2}<t_{1}$ - the positron terms. It means that the negative energy states are assumed to be not available to the electrons, the upper continuum is assumed to be not available to the positrons which are recognized as particles traveling backwards in time \cite{Wei}. As a result, the total number of degrees of freedom that is determined by the complete basis of the Dirac plane waves, is divided in half. One half of the degrees of freedom is assigned to the electron, and the other - to the positron. QED with Eq. \eqref{39} predicts the annihilation of electron-positron pairs \eqref{38} \cite{Fey,Ber,Wei}. The singlet pair of free particles with the center of mass at rest is converted with the greatest probability into two photons which, due to the momentum conservation, should be emitted exactly in the two opposite directions, at the angle 180${}^{0}$ to each other.
\par Thus, as noted in \cite{Fey}, the choice of Eq. \eqref{40} is unsatisfactory for the hole theory of the positron, but is satisfactory for the electron theory of Dirac. Further, the division of the complete basis of the Dirac plane waves into two parts leads to the following fact: neither electron states nor positron states separately form the complete system of the wave functions. 
\par Although the propagator was discussed in the work \cite{Fey}, quantum field theory with \eqref{40} has not received its development so far. Apparently, the analysis of the electron-positron system with the use of \eqref{40} was only carried out in \cite{Aga}. According to \eqref{40}, the free electron and the free positron should be treated as ordinary, different particles, each being characterized by the complete set of the Dirac plane waves. In the vacuum state the lower continua for each of these particles are completely filled and the upper continua are not occupied. Then this vacuum state is charge-neutral and stable. The latter is due to the fact that any reaction of the electrons and the positrons in the negative-energy states is forbidden by the energy conservation law since  $\delta(\varepsilon_{el}+\varepsilon_{pos}-\sum{\ }the{\ }photon{\ }energies)\equiv 0$ when both the electron energy $\varepsilon_{el}$ and the positron energy $\varepsilon_{pos}$ are negative.
\par For this completely symmetric representation, the choice of Eq. \eqref{39} is unsatisfactory, and it makes necessary to choose the propagator \eqref{40} for both the electron and the positron. In this approach, the annihilation reaction \eqref{38} is prohibited \cite{Aga}. Instead, the reaction between the electron and positron involves, together with emitted photons, the massless boson which is formed by the strongly coupled electron-positron pair. This annihilation-like process between the electron and the positron can be represented as follows:
\begin{equation}\label{41} 
e^{-}e^{+}\to B\gamma_{1}\gamma_{2},                                                     
\end{equation}
where $B$ denotes the spinless, massless composite boson, and $\gamma_{1,2}$ are the two photons emitted. Since the three particles are the products of this reaction, the $2\gamma$ angular correlation spectrum for \eqref{41} must have the fundamental width. This angular width is minimal for non-relativistic colliding beams and the center of mass at rest for the electron-positron pairs. Then the $2\gamma$ angular correlation spectrum is characterized by a narrow peak with the full-width-at-half-maximum not exceeding 0.2 mrad. In the case of the reaction \eqref{41}, the decay of the muons should be follows:
\begin{equation}\label{42}
\mu^{\pm} \to e^{\pm}B\gamma,
\end{equation}
and the other channel:
\begin{equation}\label{43}
\mu^{\pm} \to e^{\pm}B_{1}B_{2}.
\end{equation}
\par For the reaction \eqref{38} there is no fundamental width in the $2\gamma$ angular correlation spectrum. For the singlet pair of the free particles with the center of mass at rest, the $2\gamma$ correlation spectrum must be $\propto \delta(\theta-\pi)$, $\theta$ is the angle between the photons momenta.  An important circumstance is that such the correlation spectrum has never been observed. In all experiments known to us the angle $\theta \neq \pi$ and $\pi-\theta>>0.2$mrad. 
It means that the presence of the third particle on the right side of the reaction \eqref{38} is not excluded, and the reaction \eqref{41} is possible. 
\par Apparently, the first observations of the annihilation of electron–positron pairs were published in \cite{Thi,Thb,Jol,Kle,Ali}. The first detailed experimental measurements of the $2\gamma$ angular correlation spectrum were carried out in \cite{Mon}. These authors showed that the two quanta emitted in opposite directions within one degree. Then, many works were appeared in which the annihilation of positrons in condensed and gaseous matters was studied 
\cite{Deb,Dum,Mul,Pag,Lan,Hei,Pus,Gri}. In these experiments on the low-energy positron annihilation the center of mass of annihilating pairs is usually in motion with respect to an observer. Then the angle between the photon directions departs from 180$^{0}$ by an amount of the order of $v_{cm}/c$, where $v_{cm}$ is the velocity of light. Hence, the annihilation angular correlation spectra of the $2\gamma$ radiation are characterized by finite widths as well. Moreover, the published widths of $2\gamma$ angular correlation spectrum are significantly larger 0.2 mrad. So, in metals, the characteristic values of these widths are approximately equal to 4$\div$10 mrad \cite{Stu}. 
\par After that, having no doubt that in the non-relativistic case the conventional annihilation occurs, experiments began to be carried out with colliding beams of electrons and positrons with the particle energies of hundreds of keV and higher. If the $2\gamma$ spectra would be measured, then the spectral widths would be significantly larger compared to the fundamental width. The reason for this is the experimental energy-direction distributions of the particles in the beams. 
\par Thus, in the well-known experiments on the electron-positron reaction it is not possible to distinguish the reaction \eqref{41} from the conventional annihilation \eqref{38}. Possibly, the conventional annihilation takes place, but experimental evidence that the reaction products for the singlet electron-positron pairs are only two particles but definitely not three particles, was not obtained up to now. Although, as estimated in \cite{Aga}, there was an experimental setup 
\cite{Lyn}, in which a beam of non-relativistic positrons was created with parameters sufficient to answer the question which of the reactions actually takes place.
\par Thus we conclude that our treatment of the muon as the composite fermion formed by three charged electron-like leptons that are in BIC with the mass equal to 207 electron masses, does not contradict the well-known results. 
\begin{appendices} 
\section{Details of calculation for Eq. \eqref{34}} 
\renewcommand{\theequation}{A\arabic{section}}
\par From Eq. \eqref{34} we can conclude that the function $W(x,y)$ is symmetrical, $W(x,y)=W(y,x)$.  
Calculating the two integrals over $t$ on the right-hand side of \eqref{34}, we obtain:  
\setcounter{section}{1}
\begin{equation}\label{A1}   
W(x,y)=2i\alpha \Bigl( \int_{1-y}^{1} \frac{z(2-y-z)W(z,y)dz}{y(z^2+x^2)-4x(1-y-z+yz/2)}     
+\int_{1-x}^{1}\frac{z(2-x-z)W(x,z)dz}{x(z^2+y^2)-4y(1-x-z+xz/2)} \Bigr).
\end{equation} 
for the case $x,y\neq 0,1$.
According to Eqs. \eqref{27}-\eqref{28} the two-particle probability density and the normalization condition of the wave function are defined only
on the diagonal \eqref{29} that in the dimensionless values is rewritten as $x+y=1$.  In this region of the function domain Eq. \eqref{34} is reduced to:
\setcounter{section}{2}
\begin{equation}\label{A2}
W(x,1-x)=2i\alpha \Bigl( \int_{x}^{1} \frac{z(1+x-z)W(z,1-x)dz}{(1+x)(z+x)^2-2x(z^2+x^2+2x)}     
+\int_{1-x}^{1}\frac{z(2-x-z)W(x,z)dz}{x(z+x-1)^2+4(1-x)(x+z-1)} \Bigr).
\end{equation} 
\par It remains for us to define the function on the boundary of its domain of definition. From \eqref{34} we obtain:
\setcounter{section}{3}
\begin{equation}\label{A3}    
W(x,0)=\frac{i\alpha}{2}W(1,0)\frac{1}{x}\ln \frac{1+x}{1-x} +\frac{2i\alpha}{x}\int_{1-x}^{1}\frac{2-x-z}{z}W(x,z)dz,
\end{equation} 
\setcounter{section}{4}
\begin{equation}\label{A4}    
W(0,y)=\frac{i\alpha}{2}W(0,1)\frac{1}{y}\ln \frac{1+y}{1-y} +\frac{2i\alpha}{y}\int_{1-y}^{1}\frac{2-y-z}{z}W(z,y)dz,
\end{equation} 
\setcounter{section}{5}
\begin{equation}\label{A5}    
W(x,1)=i\alpha \int_{0}^{1}\frac{z(1-z)W(z,1)dz}{(z+x)^2}+2i\alpha \int_{1-x}^{1}\frac{z(2-x-z)W(x,z)dz}{x(z^2+1)-4(1-x-z+xz/2)},
\end{equation} 
\setcounter{section}{6}
\begin{equation}\label{A6}    
W(1,y)=i\alpha \int_{0}^{1}\frac{z(1-z)W(1,z)dz}{(z+y)^2}+2i\alpha \int_{1-y}^{1}\frac{z(2-y-z)W(z,y)dz}{y(z^2+1)-4(1-y-z+yz/2)}.
\end{equation} 
Finally, Eq. \eqref{28} is reduced to the form:
\setcounter{section}{7}
\begin{equation}\label{A7}   
(4\pi)^2 \int_{0}^{1}x^2 (1-x)^2 \vert W(x,1-x) \vert^2 dx=1.
\end{equation}
\par It is noteworthy that the kernel of the integral equation \eqref{34} for $W(x,y)$ is purely imaginary.
\par The procedure for solving Eqs. \eqref{A1}-\eqref{A6} with \eqref{A7} was as follows. The sought function was represented as $W = ReW  + iImW$. Then, we obtained a system of coupled integral equations for the real and imaginary parts of $W$. This system of equations was transformed into matrix form, and solved by an iterative method. In order to make sure that the resulting solution weakly depends on the matrix order, the 501x501x501 and 701x701x701 matrices were used. Before the first iteration, one of the two matrices, for definiteness, the matrix corresponding to $ImW$, was specified by an arbitrary smooth function. After $N$ iterations, both matrices of $W$ were calculated as $W=\beta W_{N-1}+(1-\beta)W_{N}$. In the case of $\beta \leq 0.9$ the integral on the left-hand side of \eqref{A7} showed damped oscillations with the iteration number. Weak convergence was observed to a value close to 1.
The values $\leq 0.95 \leq \beta \leq 0.97$ were used in the calculations. In this case, the value of the integral on the left side of  \eqref{A7} was a smooth function of the iteration number. About three hundred iterations were required for the convergence of the solution which did not depend on the initial setting of the $ImW$ matrix. 
\section{The coordinate-space wave function}
\renewcommand{\theequation}{B\arabic{section}} 
\par We found the momentum-space wave function Eq.\eqref{21} which satisfies the normalization:
\setcounter{section}{1}
\begin{equation}\label{B1}  
\int d{\bf p}_{1}\int d{\bf p}_{2} \vert F({\bf p}_{1},{\bf p}_{2})\vert^2 = 1.
\end{equation}  
The coordinate-space wave function is defined as:
\setcounter{section}{2}
\begin{equation}\label{B2}  
F(r_{1},r_{2})=\frac{1}{(2\pi)^3}\int d{\bf p}_{1} \int d{\bf p}_{2} F(p_{1},p_{2})e^{i{\bf p}_{1}{\bf r}_{1}+i{\bf p}_{2}{\bf r}_{2}}.
\end{equation}  
It easy to show that the coordinate-space wave function has the the same property:
\setcounter{section}{3}
\begin{equation}\label{B3}  
\int d{\bf r}_{1}\int d{\bf r}_{2} \vert F({\bf r}_{1},{\bf r}_{2})\vert^2=1.
\end{equation}  
Using Eq. \eqref{26}, Eq. \eqref{B3} is rewritten as:
\setcounter{section}{4}
\begin{equation}\label{B4}  
F(r_{1},r_{2})=\frac{2}{\pi r_{1}r_{2}}\lim_{\delta \to 0}\sqrt{\frac{\delta}{\pi}} \int_{0}^{E/2}p_{1}\sin(p_{1}r_{1})dp_{1} 
\int_{0}^{E/2}p_{2}\sin(p_{2}r_{2})dp_{2}\frac{Q(p_{1},p_{2})}{E-2p_{1}-2p_{2}+i\delta} 
\end{equation}  
The procedure of numerical calculations of the integrals in Eq. \eqref{B4} and, then, the limit of the numerical result is very difficult.
However we know that the function $F(r_{1},r_{2})$ is normalized. Then we can use the following method. Eq. \eqref{B4} is transformed as:
\setcounter{section}{5}
\begin{equation}\label{B5}  
F(r_{1},r_{2})=\frac{N}{r_{1}r_{2}} \int_{0}^{E/2}p_{1}\sin(p_{1}r_{1})dp_{1} 
\int_{0}^{E/2}p_{2}\sin(p_{2}r_{2})dp_{2} Q(p_{1},p_{2})\delta(E-2p_{1}-2p_{2}), 
\end{equation}  
where $N$ is the normalization constant,
\setcounter{section}{6}
\begin{equation}\label{B6}  
(4\pi)^2 \int_{0}^{E/2}p_{1}^2 dp_{1} \int_{0}^{E/2}p_{2}^2 dp_{2} \vert F(r_{1},r_{2}) \vert^2 = 1. 
\end{equation}  
\par By analogy with the above, we pass to dimensionless quantities: 
\setcounter{section}{7}
\begin{equation}\label{B7}  
u=\frac{E}{2}r_{1}, v=\frac{E}{2}r_{2}, 
\end{equation}
and introduce the dimensionless wave function: 
\setcounter{section}{8}
\begin{equation}\label{B8}  
H(u,v)=\Bigl(\frac{2}{E}\Bigr)^3 F(\frac{2}{E}u,\frac{2}{E}v)
\end{equation}  
\par Then Eq. \eqref{B5} is reduced to the form:
\setcounter{section}{9}
\begin{equation}\label{B9}  
H(u,v)= \frac{N}{uv}\int_{0}^{1} x(1-x)\sin(ux) \sin(v(1-x)) W(x,1-x)dx,
\end{equation}  
where $N$ is given by:
\setcounter{section}{10}
\begin{equation}\label{B10}  
(4\pi)^2 \int_{0}^{\infty} u^2 du \int_{0}^{\infty}v^2 dv \vert H(u,v) \vert^2=1.
\end{equation}  
\par Note that the real-space function $H(u,v)$ is determined by the values on the diagonal $x+y=1$  of the momentum-space wave function domain.
The usual characteristic of the wave function is its average radius. Taking into account the symmetry of the wave function, $H(u,v)=H(v,u)$,
the average dimensionless radius was calculated as follows:
\setcounter{section}{11}
\begin{equation}\label{B11}  
u_{av}=(4\pi)^2 \int_{0}^{\infty} u^{5/2} du \int_{0}^{\infty}v^{5/2} dv \vert H(u,v) \vert^2.
\end{equation}  
\end{appendices}

\end{document}